\documentclass{elsart}
\usepackage{graphics}
\usepackage{graphicx}
\usepackage{epsfig}
\usepackage{amssymb,amscd}
\usepackage[intlimits]{amsmath}
\usepackage{latexsym}
\usepackage{amsfonts}
\newcommand*{\be}{\begin{equation}}
\newcommand*{\ee}{\end{equation}}
\newcommand*{\la}{{\mathcal L}}
\newcommand*{\no}{\noindent}
\newcommand*{\pd}{\partial}
\newcommand*{\pdm}{\pd_{\mu}}
\newcommand*{\pdn}{\pd_{\nu}}

\newcommand*{\bea}{\begin{eqnarray}}
\newcommand*{\eea}{\end{eqnarray}}
\newcommand*{\pref}[1]{(\ref{#1})}

\newcommand*{\nn}{\nonumber}
\newcommand*{\prefr}[2]{(\ref{#1}-\ref{#2})}

\begin{document}

\begin{frontmatter}

\title{Solving a Set of Truncated Dyson-Schwinger Equations with a Globally Converging Method}

\author{Axel Maas}\ead{Axel.Maas@Physik.TU-Darmstadt.de}

\address{Gesellschaft f\"ur Schwerionenforschung mbH, Planckstr. 1, D-64291 Darmstadt, Germany}

\begin{abstract}
A globally converging numerical method to solve coupled sets of non-linear integral equations is presented. Such systems occur e.g.\ in the study of Dyson-Schwinger equations of Yang-Mills theory and QCD. The method is based on the knowledge of the qualitative properties of the solution functions in the far infrared and ultraviolet. Using this input, the full solutions are constructed using a globally convergent modified Newton iteration. Two different systems will be treated as examples: The Dyson-Schwinger equations of 3-dimensional Yang-Mills-Higgs theory provide a system of finite integrals, while those of 4-dimensional Yang-Mills theory at high temperatures are only finite after renormalization.
\end{abstract}

\end{frontmatter}

\section{Introduction}\label{Intro}

Integral equations appear in many areas of physics, and they are ubiquitous in non-perturbative field theory. Most prominent examples are the Dyson-Schwinger equations (DSEs). These equations form coupled sets of non-linear integral equations, which must be solved self-consistently to make progress beyond resummed perturbation theory. Solutions to these equations thus lie at the heart of many truly non-perturbative calculations. E.g.\ in the low-energy regime of QCD, the study of the DSEs has provided a wealth of physical insight, see e.g.\ ref.\ \cite{Alkofer:2000wg} for a review. To solve these equations, fast and reliable algorithms are desirable, which minimize the required prior knowledge.

While the theory of single linear integral equations with well-behaved integral kernels is well understood, see e.g.\ \cite{Polyanin:1999}, this is not the case for typical equations in field theories. These are normally coupled sets of non-linear equations with integrable, but singular integral kernels and not necessarily well-behaved solutions.

In many problems where solutions are only searched on a compact manifold, discretization of the integral equations allow to solve them by fix-point iteration methods. This is especially the case in non-relativistic settings, such as condensed matter problems. In fully relativistic settings, such solutions are plagued by finite volume effects and discretization errors for extremely large and extremely small momenta. This is in some cases of no importance, but in the case of non-abelian gauge theories, these regions are of special interest, as they are the origins of non-perturbative phenomena and of matching to perturbation theory. Nonetheless, methods using discretization provide a wealth of valuable informations also in these cases, see e.g.\ refs.\ \cite{Fischer:2002eq,Gruter:2004bb,torus,Fischer:2003zc}.

The aim here is to construct a globally convergent method applicable to such problems. In this paper section \ref{dse} will layout the DSEs to be solved and thus set the mathematical frame\footnote{This section includes a short description of the derivation of the equations to be solved. If the reader is exclusively interested in the numerical method, this section can be mostly skipped, except for the systems of equations to be solved. These are given in \prefr{fulleqG3d}{fulleqZ3d} for the finite system and \prefr{fulleqGft}{fulleqZft} for the renormalized system.}. The numerical method, including the implementation, will be discussed for two different systems. A finite system is obtained for 3-dimensional Yang-Mills-Higgs theory. It is treated in section \ref{numerical}. In section \ref{numerical} also an explicit example will be shown. A renormalized system is obtained for finite-temperature 4-dimensional Yang-Mills theory to be discussed in section \ref{numericalren}. In section \ref{sconcl} some concluding remarks are made. Some technical details are deferred to an appendix. A thorough discussion of the physics case can be found in refs.\ \cite{Maas:2005rf,Maas:2004se,Maas:2005hs}.

\section{Dyson-Schwinger equations}\label{dse}

The complete content of a theory is given by the corresponding Green's functions. The Dyson-Schwinger equations \cite{Dyson:1949ha} determine the Green's functions of a theory through an infinite set of coupled, non-linear integral equations. A theory is fully described by the Lagrangian density $\la(\phi^a,\pd_\mu \phi^a,...)$ of the fields $\phi^a$ and their derivatives, where $a$ is a generic multi-index. The Green's functions\footnote{The discussion here is restricted to an Euclidean space-time, but can directly be transferred to any other metric.} are then given by functional derivatives w.r.t.\ the fields $\phi^b$ of the identity
\be
\int{\mathcal D}\phi\frac{\delta}{\delta\phi^a}\exp(-\int d^dx\la)=0.\label{dses}
\ee
\no Here $d$ is the number of space-time dimensions. See refs.\ \cite{Alkofer:2000wg,Rivers:1987hi} for a detailed introduction to DSEs.

It is in general not possible to solve all DSEs simultaneously, and it is therefore necessary to truncate the system to a finite number of equations. This entails nearly always violations of internal symmetries such as gauge symmetries, and it is an essentially non-trivial task to compensate or estimate these effects. For a detailed study of this problem, see e.g.\ refs.\ \cite{Alkofer:2000wg,Fischer:2003zc,Maas:2005rf}. The consequences of these truncation and how to control them is a physical issue, which will not be treated here. The implementation of appropriate measures to deal with this problem can have impact on the numerical method. Furthermore, in quantum field theory, terms in the DSEs usually diverge, and must be made finite by renormalization \cite{Bohm:yx}. Although this is again a physical issue, it has also consequences for the numerical method. Therefore some numerical aspects of both problems will be discussed briefly in section \ref{numericalren}.

The presented numerical method has been applied successfully to two systems. One is 3-dimensional Yang-Mills theory coupled to an adjoint Higgs as the infinite-temperature limit of 4-dimensional Yang-Mills theory \cite{Maas:2005rf,Maas:2004se}. The second is 4-dimensional Yang-Mills theory at high temperature \cite{Maas:2005rf,Maas:2005hs}. The first system is finite and the numerical problem thus will be treated first in section \ref{numerical}. The second requires renormalization and the necessary alterations will be discussed in section \ref{numericalren}. In the following subsections, the Dyson-Schwinger equations of each system will be shortly introduced.

\subsection{3-dimensional Yang-Mills theory coupled to an adjoint, massive Higgs}

The infinite-temperature limit of 4-dimensional Yang-Mills theory is described by a 3-dimensional Yang-Mills theory coupled to an adjoint, massive Higgs field \cite{Kajantie:1995dw}. For physical reasons the choice of Landau gauge is advantageous, which will be made here exclusively. The Lagrangian of this system in Euclidean space is then given by \cite{Maas:2004se,Kajantie:1995dw}
\be
{\mathcal L}=\frac{1}{4}F_{\mu\nu}^aF_{\mu\nu}^a+\bar c^a \pdm D_\mu^{ab} c^b+\frac{1}{2}(D_\mu^{ab}\phi^b D_\mu^{ac}\phi^c+m_h^2\phi^a\phi^a)+\frac{h}{4}\phi^a\phi^a\phi^b\phi^b,\label{lagrange}
\ee
with the field strength tensor $F_{\mu\nu}^a$ and the covariant derivative $D_\mu^{ab}$ defined as
\bea
F^a_{\mu\nu}&=&\pdm A_\nu^a-\pdn A_\mu^a-g_3f^{abc}A_\mu^bA_\nu^c\label{fstrength}\\
D_\mu^{ab}&=&\delta^{ab}\pdm+g_3f^{abc}A_\mu^c.\label{coderiv}
\eea
\noindent $A_\mu^a$ is the 3-dimensional gluon field, $c^a$ and $\bar c^a$ are the Faddeev-Popov ghost and anti-ghost fields, $\phi^a$ is the Higgs field, $g_3$ is the dimensionful coupling, which is 1 in appropriate units, and $m_h\sim g_3^2$ is fixed by Monte Carlo calculations on discrete space-time lattices to be $0.88g_3^2$ \cite{Cucchieri:2001tw}. The constants $f^{abc}$ are the structure constants of the gauge group. In general, contributions stemming from the Higgs field could be divergent. However, this is not permissible when viewing the Lagrangian \pref{lagrange} as the infinite-temperature limit of 4-dimensional Yang-Mills theory. Therefore leading order perturbation theory requires that $h$ is fixed to
\be
h=-2g_3^2\frac{C_A}{C_A+2},\label{hvalue}
\ee
\noindent where $C_A=f_{abc}f^{abc}$ is the adjoint Casimir of the gauge group.

The simplest non-vanishing Green's functions in this theory are obtained by deriving the identity \pref{dses} once more with respect to the fields. The inverse of these Green's functions are called propagators, and there is one for each field. These propagators $D_G$, $D_{\mu\nu}$, and $D_H$ of the ghost, the gluon, and the Higgs, respectively,  can be parameterized by three scalar functions, so-called dressing functions, $G$, $Z$, and $H$ as
\bea
D_G(p^2)&=&\frac{-G(p^2)}{p^2}\\
D_{\mu\nu}(p)&=&\left(\delta_{\mu\nu}-\frac{p_\mu p_\nu}{p^2}\right)\frac{Z(p^2)}{p^2}\\
D_H(p^2)&=&\frac{H(p^2)}{p^2},
\eea
\noindent respectively. The dressing functions $G$, $Z$, and $H$ have to be positive semi-definite by the Gribov condition \cite{Maas:2004se,Gribov:1977wm}. Using these dimensionless quantities instead of the propagators improves the numerical stability for large $p$ significantly, as otherwise trivial kinematic effects would require an enormous precision.

The truncated DSEs are then obtained by standard techniques \cite{Alkofer:2000wg} in a straight-forward but tedious way \cite{Maas:2005rf}. The tensor equation for the gluon propagator is contracted with an appropriate projector to yield a scalar equation. The projector is parameterized by a variable $\zeta$, on which therefore the integral kernels depend. This yields the DSEs for the ghost dressing function $G$, the gluon dressing function $Z$, and the Higgs dressing function $H$
\bea
0&=&1+\frac{g_3^2C_A}{(2\pi)^2}\int d\theta dq A_T(p,q)G(q)Z(p-q)-\frac{1}{G(p)}\label{fulleqG3d}\\
0&=&1+\frac{m_h^2}{p^2}+T^{HG}+T^{HH}+\frac{g_3^2C_A}{(2\pi)^2}\int d\theta dq\Big(N_1(p,q)H(q)Z(p+q)\nonumber\\
&&+N_2(p,q)H(p+q)Z(q)\Big)-\frac{1}{H(p)}\label{fulleqH3d}\\
0&=&1+T^{GH}+T^{GG}+\frac{g_3^2C_A}{(2\pi)^2}\int d\theta dq\Big(R(p,q)G(q)G(p+q)\nonumber\\
&&+M_L(p,q)H(q)H(p+q)+M_T(p,q)Z(q)Z(p+q)\Big)-\frac{1}{Z(p)},\label{fulleqZ3d}
\eea
\noindent where $\theta$ is the angle between $p$ and $q$. The tadpoles $T^{ij}$ are not independent functions but compensate any divergencies and thus render the equations \prefr{fulleqG3d}{fulleqZ3d} finite \cite{Maas:2005rf}. The integral kernels\footnote{The kernel $M_T$ in addition depends on a parameter $\delta_{3g}$. It is fixed throughout to $1/4$ and parameterizes the constructed three-gluon vertex. Furthermore this induces an additional non-trivial dependence of $M_T$ on $G$ and $Z$, see appendix \ref{appkernels} and ref.\ \cite{Maas:2004se}.} $A_T$, $N_1$, $N_2$, $R$, $M_T$, and $M_L$ as well as the tadpoles are listed in appendix \ref{appkernels}. Trivial factors, such as the integral measure, have been included into the kernels. 
The kernels all have a very similar structure, being rational functions of the momenta. All these kernels furthermore contain integrable singularities of the type $1/(p-q)^m$ with some positive $m$.

\subsection{Finite temperature Yang-Mills theory}

The second example is finite temperature 4-dimensional Yang-Mills theory. In the Matsubara formalism \cite{Kapusta:tk} it is described by the (Euclidean) Lagrangian
\be
{\mathcal L}=\frac{1}{4}F_{\mu\nu}^aF_{\mu\nu}^a+\bar c^a \pdm D_\mu^{ab} c^b,\label{lagrange2}
\ee
\no which is very similar to \pref{lagrange}. The only difference to the previous example is replacing $g_3$ by the 4-dimensional dimensionless coupling $g_4$ in the field strength tensor \pref{fstrength} and the covariant derivative \pref{coderiv} and dropping the Higgs field. Using the Matsubara formalism \cite{Kapusta:tk} the DSEs of the vacuum \cite{Alkofer:2000wg} can be extended to finite temperature \cite{Maas:2005rf}. The propagators are again the most simple Green's functions and can be parameterized by also three functions
\be
D_G(p)=\frac{-G(p_0^2,\vec p^2)}{p^2}\label{ghostDressing}
\ee
for the ghost propagator and similarly for the gluon propagator as \cite{Kapusta:tk}
\be
D_{\mu\nu}(p)=P_{T\mu\nu}(p)\frac{Z(p_0^2,\vec p^2)}{p^2}+P_{L\mu\nu}(p)\frac{H(p_0^2,\vec p^2)}{p^2}.\label{gluonDressing}
\ee
where $P_T$ and $P_L$ are projectors transverse and longitudinal w.r.t.\ the heat-bath \cite{Kapusta:tk}. The coincidence of the names of the dressing functions to those of the previous example is intentional. There is an exact correspondence in the limit of infinite temperature.

The truncated DSEs for the dressing functions $G$, $Z$, and $H$ are obtained in the same way as before and read
\bea
0&=&\widetilde{Z}_3+\frac{g_4^2TC_A}{(2\pi)^2}\sum_{n=-\infty}^{\infty}\int d\theta dq \Big(A^f_T(p,q)G(q)Z(p-q)\nonumber\\
&&+A^f_L(p,q)G(q)H(p-q)\Big)-\frac{1}{G(p)}\label{fulleqGft}\\
0&=&\xi Z_{3L}+T^{fHG}+T^{fHH}+\frac{g_4^2TC_A}{(2\pi)^2}\sum_{n=-\infty}^{\infty}\int d\theta dq\Big(P^f(p,q)G(q)G(p+q)\nonumber\\
&&+N^f_L(p,q)Z(q)Z(p+q)+N^f_1(p,q)H(q)Z(p+q)\nonumber\\
&&+N^f_2(p,q)H(p+q)Z(q)+N^f_T(p,q)H(q)H(p+q)\Big)-\frac{\xi}{H(p)}\label{fulleqHft}\\
0&=&Z_{3T}+T^{fGH}+T^{fGG}+\frac{g_4^2TC_A}{(2\pi)^2}\sum_{n=-\infty}^{\infty}\int d\theta dq\Big(R^f(p,q)G(q)G(p+q)\nonumber\\
&&+M^f_L(p,q)H(q)H(p+q)+M^f_1(p,q)H(q)Z(p+q)\nonumber\\
&&+M^f_2(p,q)H(p+q)Z(q)+M^f_T(p,q)Z(q)Z(p+q)\Big)\nonumber\\
&&+\frac{p_0^2(\zeta-1)}{2p^2}\left(Z_{3L}-\frac{1}{H(p)}\right)-\frac{1}{Z(p)}.\label{fulleqZft}
\eea
\noindent The temperature is denoted by $T$. The summation is performed over all integers $n$. In the Matsubara formalism the component $q_0$ is discrete and given by $q_0=2\pi T n$, called the Matsubara frequency. Also the external $p_0$ is thus discrete.

In the present case, the sum is truncated, and only the range $[-N+1,N-1]$ is included. Close inspection of the equations reveals further that the dressing functions can only depend on $\left|p_0\right|$ and $\left|\vec p\right|$. The corresponding symmetry, under $p_0\to -p_0$, is used to reduce the number of equations significantly. Therefore $3N$ independent functions have to be determined. The largest system treated numerically so far is $N=21$.

The kernels $A^f_T$, $A^f_L$, $R^f$, $M^f_T$, $M^f_1$, $M^f_2$, $M^f_L$, $P^f$, $N^f_T$, $N^f_1$, $N^f_2$, and $N^f_L$ are rather lengthy, and will not be displayed here. They can be found in ref.\ \cite{Maas:2005rf}. The kernels are very similar in structure to their 3-dimensional pendants listed in appendix \ref{appkernels}. The same applies to the tadpoles $T^{fij}$, with the notable exception of th $T^{fHj}$. These are much more similar to the transverse expressions $T^{fGj}$ then in the case of 3 dimensions. Especially these are now also determined by integrals, see ref.\ \cite{Maas:2005rf} again for details. The finite-temperature theory is renormalizable. Thus explicit wave-function renormalization constants $\widetilde{Z}_3$, $Z_{3L}$, and $Z_{3T}$ have been introduced\footnote{No vertex renormalization occurs within this truncated system.}.

The variables $\zeta$ and $\xi$ are introduced by contraction of the equation of the gluon propagator with two different projectors \cite{Maas:2005rf,Maas:2005hs} as before. These variables have been introduced to study the consequences of the truncation. For $p_0=0$, equation \pref{fulleqHft} is only superficially dependent on $\xi$: since all integral kernels are proportional to $\xi$, the dependence can be divided out for $\xi\neq 0$.

The renormalization constants are defined at a subtraction point $s$, which can be chosen arbitrarily, but for numerical reasons should be different from 0. The explicit implementation of the renormalization prescription \cite{Maas:2005rf,Maas:2005hs} for the DSE of a dressing function $F=G$, $Z$, $H$ with self-energy contributions $I$
\be
\frac{1}{F(p)}=1+I(p)\label{ren1}
\ee
\noindent is then for the corresponding wave function renormalization constant $Z_3$
\bea
\frac{1}{F(p)}&=&1+\delta Z_3+I(p)\\
\delta Z_3&=&-I(s)\\
Z_3&=&1+\delta Z_3
\eea
\noindent and in the case of $H(0,|\vec p|)$
\bea
\frac{1}{H(0,|\vec p|)}&=&1+\delta Z_{3L}+\frac{\delta m^2}{p^2}+I(p)\\
\delta m^2&=&m_r^2-\lim_{p\to 2\delta}p^2I(p)\\
\delta Z_{3L}&=&-I(s)+\frac{\lim_{p\to 2\delta}p^2I(p)}{s^2},\label{ren2}
\eea
\noindent where $m_r=m_{3d}$ is the renormalized mass, and $\delta$ is an infinitesimal parameter, different from zero only for numerical reasons. It is taken to be the same value as the infrared integral cutoff introduced in the next section. The difference compared to performing the subtraction at 0 is negligible. Note that due to the truncation of the Matsubara sum the wave function renormalization constants depend on $p_0$. This dependence vanishes for $N\to\infty$ \cite{Maas:2005hs}, but has to be taken into account in the numerical treatment presented here.

\section{Numerical method}\label{numerical}

The numerical technique to solve the system of equations \prefr{fulleqG3d}{fulleqZ3d} and \prefr{fulleqGft}{fulleqZft} are very similar. Renormalization in the 4-dimensional system \prefr{fulleqGft}{fulleqZft} requires some extensions. Therefore here only the finite system will be treated. The modifications for the renormalized system will be discussed in section \ref{numericalren}.

The earliest approach to solve the 4-dimensional equivalent (without the additional Higgs field) of system \prefr{fulleqG3d}{fulleqZ3d} was based on a direct iteration method \cite{Hauck:1996sm}. This has afterwards been replaced by a method based on a local Newton iteration \cite{Fischer:2003zc,Atkinson:1997tu,Fischer:2002hn}. The latter approach used as starting inputs for a local iteration results obtained from a discretized version of the problem \cite{Fischer:2002eq}. The method presented here generalizes this ansatz to provide global convergence and remove the necessity of input data. For the sake of completeness the full method will be described here, although several elements have been taken from the previous one.

The basis for the numerical treatment is an analytical investigation of the asymptotic behavior of the solutions. This will be discussed in subsection \ref{ssanalytic}. An important ingredient is the representation of the solutions using their asymptotics. This and the integration method will be discussed in subsection \ref{ssexpint}. The heart of the method is the iteration procedure, which will be described in subsection \ref{ssheart}. Some notes on the efficient implementation are given in subsection \ref{ssopti}. An explicit calculation with a typical set of numerical parameters is given as an example in section \ref{ssexample}.

\subsection{Analytical foundation}\label{ssanalytic}

It is important to gather as many analytical properties as possible to obtain an appropriate representation of the solutions. Especially using as few parameters as possible permits to obtain maximal numerical efficiency.

The searched for positive semi-definite solution functions are continuously differentiable and analytical in the momentum range $(0,\infty)$. For physical reasons no poles at intermediate momenta are expected. However the solutions are not bounded from above. Thus divergences at $p=0$ are possible. Therefore the infrared asymptotic behavior is of special interest. Favorable, it turns out that the leading asymptotic behavior can be obtained analytically \cite{Maas:2005rf,Maas:2004se,Zwanziger:2001kw}.

In the infrared, power-like ans\"atze can be used \cite{Maas:2004se,Zwanziger:2001kw}\footnote{Note that this is very similar to the case of 4 dimensions \cite{vonSmekal:1997is}.},
\be
\lim_{p\to 0}F(p)=A_F(p^2)^{-e_F}.\label{infrared}
\ee
\noindent This is possible due to the singularity structure $1/(p-q)^m$ of the integral kernels, which lead to a domination of the integrals by momenta of order $q\approx p$. It is hence allowed to replace the dressing functions by the ans\"atze \pref{infrared}. It is then possible to solve the massless integrals analytically \cite{Maas:2004se,Zwanziger:2001kw}. The Higgs propagator is dominated by the renormalized mass term in equation \pref{fulleqH3d} and thus has the unique exponent $e_H=-1$. This also determines $A_H$ uniquely \cite{Maas:2004se}. There are two possible solutions for the ghost exponents. One of the solution is $e_G=1/2$, while the other depends on $\zeta$ and varies as $\e_G\in(1/4,1/2]$. The gluon exponent $e_Z$ is not independent, but related to $e_G$ by \cite{Maas:2004se,Zwanziger:2001kw}
\be
e_G=-\frac{1}{2}\left(e_Z+\frac{1}{2}\right).
\ee
\noindent The coefficients $A_G$ and $A_Z$ cannot be determined uniquely. However, they are related by \cite{Maas:2004se}
\bea
\frac{1}{A_G^2 A_Z}&=&\frac{C_Ag_3^2}{(4\pi)^\frac{3}{2}}\frac{2^{4(e_G-1)}(2+2e_G(\zeta-1)-\zeta)\Gamma(2-2e_G)\sin^2(\pi e_G)}{\cos(2\pi e_G)(e_G-1)e_G^2\Gamma(\frac{3}{2}-2e_G)}\label{agaz}\\
&\equiv&\frac{1}{J(e_G,\zeta)}\nonumber.
\eea

Furthermore, in the ultraviolet, $p\gg g_3^2$, an analytic solution can be obtained by perturbation theory due to asymptotic freedom \cite{Maas:2004se}. The dressing functions are then given by
\be
\lim_{p\to\infty}F(p)=1+\frac{w_Fg_3^2C_A}{p},\label{perturbative}
\ee
\noindent The $w_F$ are the positive constants $w_G=1/16$, $w_Z=9/64$ and $w_H=1/4$. Note that in contrast to 4 dimensions these are already the resummed solutions, as leading order and leading order resummed perturbation theory in 3-dimensional Yang-Mills theory are the same \cite{Maas:2004se}. This concludes the list of known analytical properties of the solutions.

\subsection{Expansion and integration}\label{ssexpint}

The first step is to describe the solutions in terms of known functions. The solutions are due to their ultraviolet behavior neither  integrable nor square-integrable. This prohibits an expansion in any orthogonal set of polynomials. This problem is reinforced by the infrared divergence of the ghost dressing function.

Therefore originally \cite{Atkinson:1997tu,Bloch:1995dd} the dressing functions were expanded as
\bea
F(p)&=&\theta(\varepsilon-p)D_{F}^I(p)\nonumber\\
&&+\theta(p-\varepsilon)\theta(\Lambda-p)D_{F}^N(p)+\theta(p-\Lambda)D_{F}^U(p)\label{fit}\\
\ln\left(D_{i}^N(p)\right)&=&\sum_{j=0}^{N_{ch}}c_{ij}T_j(M(p))\label{chebyexp}
\eea
\noindent where $D^I_F$ denotes the analytical infrared solution \pref{infrared} and $D^U_F$ the analytical ultraviolet solution \pref{perturbative}. The index $i$ runs through 1, 2, 3 corresponding to $G$, $Z$, and $H$, respectively. The range $[\varepsilon,\Lambda]$ is covered by a numerical expansion\footnote{Expanding the logarithm instead of the function itself leads to a considerable smoothing at large momenta and the number of needed Chebychef polynomials drastically decreases.} in $N_{ch}$ Chebychef polynomials $T_j$. Good results can be achieved with $N_{ch}=30-40$. For special purposes like computation of Schwinger functions or thermodynamic quantities \cite{Maas:2004se} $N_{ch}=100-500$ are used.

The typical expansion range in the case of 3 dimensions is chosen to be $[\varepsilon,\Lambda]=[10^{-2}g_3^2,2\cdot 10^3 g_3^2]$. The function $M$ in \pref{chebyexp} is a suitable mapping function which maps the momenta to the domain  $(-1,1)$ of the Chebychef polynomials. Since the dressing functions $F=G$, $Z$, $H$ vary on logarithmic scales, the mapping function is chosen accordingly as\footnote{An alternative is a conformal mapping like $z/(z+1)$ which removes the necessity of a numerical upper and lower cutoff of the integrals. However, such a mapping does not provide any advantage beyond formal elegance and in addition did not provide as well spaced evaluation points as the logarithmic mapping.}
\be
M(p)=A+B\ln(p)\label{map}
\ee
\noindent where $A$ and $B$ are chosen such as to yield
\bea
M^{-1}(\varepsilon)&=&z_{0}\\
M^{-1}(\Lambda)&=&z_{N_{ch}},
\eea
\noindent where $M^{-1}$ is the inverse function of \pref{map} and $z_i$ are the $i$th zero of the Chebychef polynom of order $N_{ch}$.

To increase the numerical stability, \pref{fit} is altered to
\bea
D_F(p)&=&\theta(\varepsilon-p)D_{F}^I(p)\nonumber\\
&+&\theta(p-\varepsilon)\theta(\Lambda-p)D_{F}^N(p)f_F(p)+\theta(p-\Lambda)D_{F}^U(p)\label{ffit}
\eea
\noindent where the $f_F$ interpolate between the qualitative behavior of $D_{F}^I$ and $D_{F}^U$. Thus for divergent quantities like the ghost dressing function
\be
f_G(p)=1+\frac{a_G}{p^{-2e_G}}\frac{b_G+p^{-2e_G}}{c_G+p}\label{gfit}
\ee
\noindent with suitably chosen fitting constants $a_G$, $b_G$, $c_G$.  This especially improved the quality at large $p$, as it is notoriously difficult to fit with a polynomial onto a constant. Similar for converging quantities as the gluon and the Higgs the fits
\bea
f_Z(p)&=&\frac{p^{-2e_Z}}{c_Z+p^{-2e_Z}}\left(1+\frac{a_Z}{b_Z+p}\right)\label{zfit}\\
f_H(p)&=&\frac{a_Hp^{-2e_H}}{p^{-2e_H}+c_Hp+b_H^2}\label{hfit}
\eea
\noindent are used, respectively. Note that it turns out to be more stable to neglect in the case of the Higgs the subleading contributions in \pref{perturbative} fitted by $c_H$. The coefficients are chosen to resemble the typical scale $\nu$ for the system investigated. In case of the 3-dimensional theory this is $\nu=g_3^2$ while in case of the 4-dimensional system $\nu=g_4^2T$. The coefficients can then be read off table \ref{tabcoeff} for the 3-dimensional theory. In the 4-dimensional theory a more simplified version of \prefr{gfit}{hfit} is used, omitting the perturbative tail. For the hard modes, $p_0\neq 0$, essentially a fit of type \pref{hfit} is used.

\begin{table}
\begin{center}
\begin{tabular}{|c|c|c|c|}
\hline
Dressing function ($F$) & $a_F$ & $b_F$ & $c_F$ \cr
\hline
Ghost ($G$) & 1 & $\nu^{2e_G}$ & $\nu$ \cr
\hline
Gluon ($Z$) & $\nu$ & $\nu$ & $2/\nu^{2e_Z}$ \cr
\hline
Higgs ($H$) & 1 & $m_r^2$ & 0 \cr
\hline
\end{tabular}
\end{center}
\caption{The coefficients for the fitting functions \prefr{gfit}{hfit} in terms of the scale of the problem $\nu$ and of the renormalized Higgs mass $m_r$ \cite{Maas:2004se}.}
\label{tabcoeff}
\end{table}

Additionally it is necessary to choose an integrator. For the radial integral, after applying the logarithmic map \pref{map} to the integration variable, Gauss-Legendre integration is sufficient \cite{Bloch:1995dd}. Since the integrals are finite, a lower and upper numerical integral cutoff has been used, and the integration has been performed in $[\delta,\Omega]$, with $\delta\ll\varepsilon$ and $\Lambda\ll\Omega$, typically $[10^{-5}g_3^2,10^7g_3^2]$. The remaining problem to be dealt with is the singularity structure of the integral kernels and of the ghost dressing function $G$. These appear at $|q|=0$ and $p=\pm q$, depending on the equation. The $|q|=0$ singularity is cured by imposing the lower cutoff. Here the logarithmic mapping of the integration variable provides a sufficient sampling of the integrable singularities arising. More problematic is the singularity due to the coincidence with the external momentum. It can be removed by splitting the integral in two parts, up to and above $p$. Again the logarithmic spreading provides sufficient sampling of the integrable singularity. As $p$ will only be chosen from $[\varepsilon,\Lambda]$  the integral is finally performed as
\be
\int_0^\infty\to\int_\delta^\varepsilon+\int_\varepsilon^p+\int_p^\Lambda+\int_\Lambda^\Omega.
\ee
\noindent The angular integral is straightforward and can be done by normal Gauss-Legendre integration. However, it is necessary to perform the angular integral before the radial one, as some divergences cancel only due to the angular integration. These cancellations also require a sufficient sampling, leading to $N_a=80$ points in angular direction and $N_\delta+N_\varepsilon+N_p+N_\Lambda$$=60+100+100+60=320$ points in radial direction as typical values.

\subsection{Micro-, Macro, and Super-cycles}\label{ssheart}

Originally, solutions for the set of coefficients $c_{ij}$ were obtained using a local Newton iteration procedure \cite{Atkinson:1997tu}. The presented method here uses instead a global Newton method with back-tracking \cite{Kelley:1995}. This approach is very insensitive to the starting value, and a typical starting guess is $c_{ij}=0$. The necessary $3N_{ch}$ functions to determine the coefficients $c_{ij}$ are obtained by evaluating the system of equations \prefr{fulleqG3d}{fulleqZ3d} at the zeros of $T_{N_{ch}}$. At these points the expansion in Chebychefs is exact. This also entails that all evaluation points stem from $[\varepsilon,\Lambda]$. This makes it even more necessary to know the behavior of the functions outside this domain analytically to perform the integrals in the intervals $[\delta,\varepsilon]$ and $[\Lambda,\Omega]$.

Therefore the numerical problem was to solve the $3N_{ch}$ non-linear equations for the coefficients $c_{ij}$ and determine the coefficient $A_G$. These equations can be collected generically into an equation vector $\vec E$, with components defined as
\be
E_{(i-1)N_{ch}+j}=\tau(p_j)+\sum_{l=1}^3 I(F_i,F_l,p_j)-\frac{1}{F_i(p_j)} \label{beqset}
\ee
\noindent with $F_1=G$, $F_2=Z$, and $F_3=H$ and $i=1$, 2, 3. The index $j$ runs from 1 to $N_{ch}$, $I$ are the integrals, and $\tau$ the remaining tree-level terms. Of course, an exact solution of the system satisfies $\vec E=\vec 0$, and thus the aim is to find such a solution.

New values $c_{ij}'$ for the coefficients are then obtained iteratively by \cite{Kelley:1995}
\be
\vec c'=\vec c-\lambda J^{-1} \vec c,\label{cchange}
\ee
\no where $\vec c$ is the vector of coefficients $c_{ij}$ with the components $(\vec c)_{(i-1)N_{ch}+j}=c_{ij}$. $J$ is the Jacobian defined as
\be
J_{ij}=\frac{\pd E_{i}}{\pd c_{j}}.
\ee
\no The stepwidth is determined by the backtracking parameter $\lambda$. It is chosen such as to minimize $|\vec E|$, where a simple quadratic interpolation has been used \cite{Kelley:1995}: A first trial step is a full local Newton step ($\lambda=1$). This is accepted, if it decreases the previous value $|\vec E|=r_0$. If this is the case, the algorithm has already moved so close to the correct solution that the faster local algorithm can be used (which provides quadratic instead of the linear convergence of the global version \cite{Kelley:1995}).

If this not leads to a decrease, a trial step of size $\sigma_1<1$ is performed. If this again provides no reduction of $|\vec E|$, the quadratic interpolation is used: Given the step-size of the current trial step $\lambda_2$ and of the previous trial step $\lambda_1$, leading to values $r_2$ and $r_1$ of $|\vec E|$, respectively, a new trial value of $\lambda$ is determined by a quadratic interpolation using the first two terms of a Taylor expansion of $|\vec E|$ to construct \cite{Kelley:1995}
\bea
p_1&=&\frac{r_0(\lambda_2^2-\lambda_1^2)-\lambda_1^2 r_2+\lambda_2^2 r_1}{\lambda_1\lambda_2^2-\lambda_1^2\lambda_2}\nn\\
p_2&=&\frac{(2\lambda_1\lambda_2)(\lambda_2(r_1-r_0)-\lambda_1(r_2-r_0))}{\lambda_1-\lambda_2}\nn.
\eea
\no Thus if $p_2$ is greater than zero a descent direction is found, and the new value $\lambda=-p_1/p_2$ minimizes the Taylor series truncated at the quadratic order. However, the quadratic approximation may not be sufficient, leading to a step-width outside the interval $(0,\lambda_2]$ or may be too close to $0$ or $\lambda_2$ to be reliable. Thus if the new $\lambda$ lies outside the range $[\sigma_0\lambda_2,\sigma_1\lambda_2]$, with $\sigma_0<\sigma_1$, the new value of $\lambda$ is taken to be the corresponding border value, e.\ g.\ if the new value of $\lambda$ would be larger than $\sigma_1\lambda_2$ it is reduced to $\sigma_1\lambda_2$. If $p_2$ is positive or zero, a quadratic approximation is not reliable, and the new trial step is selected as $\lambda=\sigma_1\lambda_2$.With this new value of $\lambda$, a new trial step is performed. If still no decrease is found, this procedure is repeated until $|\vec E|$ is decreased, and the trial step is accepted and \pref{cchange} is performed. This is repeated until convergence is achived, which will be discussed below.

The method fails to converge if $\lambda$ becomes less than a predetermined value $\alpha_G$. In the present case $\alpha_G$ is essentially determined by empirical knowledge, but in general cases should be at least of the order of the machine precision. The values of $\sigma_0$ and $\sigma_1$ are tuning parameters for the algorithm, a specific choice is presented in subsection \ref{ssexample}.

A significant problem\footnote{Special thanks to Claus Feuchter and Burghard Gr\"uter for a very inspiring discussion on this topic.} is the determination of the coefficient $A_G$ of the infrared asymptotic behavior  \pref{infrared} of the ghost dressing function $G$. Trying to incorporate it into the global Newton method failed. Hence this part is moved to an outside iteration: after each global Newton iteration, a new value of $A_G$ is determined by requiring that the infrafred solution $D_G^I$ connects continuously to the intermediate function $f_GD_G^N$. In general this leads to a continuous but not to a continuously differentiable function until convergence has been achieved.

Therefore $A_G$ has not yet its final value. Thus the condition \pref{agaz} is not fulfilled. Hence to stabilize the system, $A_Z$ is not determined by its functional relation to $A_G$, but left as a free parameter and determined in the same way as $A_G$. Even for $A_H$, where its final value is exactly known, it is more advantageous to let it as a free parameter to stabilize the numerical algorithm. The starting values for $A_G$, $A_Z$, and $A_H$ are chosen such that the infrared solution and the intermediate momentum solution fit continuously after the initial $c_{ij}$ have been set. After each global Newton run, the coefficients are refitted. Thus the Newton iteration is termed a micro-cycle while the fix-point iteration of the coefficients is termed macro-cycle.

This approach is feasible, as it turns out that the radius of convergence for the coefficients is significantly larger than for the Chebychef coefficients. $A_G$ is strongly constrained by the vanishing of the inverse ghost dressing function in the infrared, thus determining $A_Z$ and $A_H$ by their exact solutions. In other cases possibly another global iteration for the coefficients would be necessary. Note that during the iteration the values of the $A_F$ can change by several orders of magnitude, even if initial and final value are close by. Furthermore as \pref{agaz} is not fulfilled with the initial values it makes no sense to run the global Newton method until it converges. Therefore after a number of steps $N_N$ the Newton method is interrupted, if it did not yet converge, to make a new macro-cycle step and then start a new micro-cycle. After several macro-cycle steps the Newton method already converges, i.e. $|\vec E|<\delta_G$ for a fixed $\delta_G$, in less than $N_N$ steps. In the present case it was possible to choose $N_N$ fixed. In more instable solutions an adaptive choice may be appropiate.

In addition, still slow convergence is encountered. To deal with it leads to the introduction of super-cycles: it turns out that convergence is extremely slow if $\varepsilon$ is chosen already quite small from the beginning. In this case the presented initial values are far from the final ones, and only very small steps can be made \cite{Kelley:1995}. Since the infrared regime and the finite momenta regime do separate in the system under investigation, it is possible to choose an $\varepsilon$ larger than the final matching scale. The solutions then are not continuous differentiable at $\varepsilon$ and the relation \pref{agaz} is not fulfilled. Thus the independence of the coefficients $A_F$ in the fitting procedure is important here as well. Using the previous solutions for a given $\varepsilon$ as the input for a smaller one, it was possible to approach the correct solutions stepwise with sufficiently good convergence. Indeed, regularly only the first steps of the micro-cycle in the first one or two macro-cycle steps utilized the global properties of the method and afterwards local and thus quadratic convergence in $|\vec E|$ was achieved. This is repeated until \pref{agaz} is satisfied to the required precision.

It is possible to reduce the number of necessary super-cycle steps drastically in two ways. One is an improvement of the fit functions $D_F^U$ and $f_F$ to include sub-leading corrections. This leads to the final form of equation \pref{ffit} and the fits \prefr{gfit}{hfit}. The other is the introduction of damping in the determination of the $A_F$. To this end the new values $A_F'$ providing a continuous fit between $D_F^I$ and $f_FD_F^N$ are computed, but instead setting the value of $A_F''$ in the next step to $A_F'$, only a variable admixture of $A_F'$ to the original $A_F$ is chosen as
\be
A_F''=\frac{dA_F+A_F'}{d+1},
\ee
\no where $d\ge 0$ is the damping parameter.

To test whether the algorithm has succeeded, two different criterions for the micro- and macro-cycle are used. For a micro-cycle the absolute value of $|\vec E|$ is used. The typical scale of terms is set by the tree-level term of order 1. Compared to this scale a mean deviation of the order $10^{-8}$ is achieved with the described numerical parameters. The largest deviation are usually found in the far infrared of equation \pref{fulleqZ3d}, where diverging terms have to cancel and thus very high accuracy is necessary.

If the available precision is not sufficient, the necessity for these cancellations is also a limiting factor for the algorithm. An example of this problem is encountered when $\zeta\approx 3$ in case of the varying solution branch described in subsection \ref{ssanalytic}. In this case, the divergencies of the two competing terms are softened, and thus very small values of $\varepsilon$ are necessary before only the infrared leading terms are relevant and thus the asymptotic infrared solution $D_F^I$ is sufficient. If $\zeta$ is sufficently close to 3, the required values of $\varepsilon$ are so small that the cancellations cannot be resolved with the available precision and the algorithm fails to converge.

For the macro-cycle the largest rate of change of the $A_F$ is used as a measure of convergence. Typical when the rate of change is below a fixed value $\delta_I$ of the order of a per mill the macro-cycle is completed. For the super-cycle the ratio $A_G^2A_Z/J(\zeta,e_G)$ according to \pref{agaz} is used\footnote{The value of $A_H$ is always so close to the final value that this is not a relevant figure of merit.}. Note that due to numerical errors it is even with a machine precision of $10^{-16}$ only possible to fulfill \pref{agaz} to the order of $10^{-5}$.

\subsection{Notes on implementation and optimization}\label{ssopti}

While the implementation of the macro- and super-cycles is straightforward, there are some subtleties concerning the implementation of the micro-cycle. Once the values of the $A_F$ are fixed in the macro-cycle, calculating the current value of the right hand sides of the DSEs is straightforward numerically. However, as the evaluation of the exponentials in the fits of the $F$ is numerically costly, it is much more efficient to calculate all equations for a given $p_j$ simultaneously, as here the same set of Gauss-Legendre points can be used.

This issue is much more important when performing the Newton step. To determine the descent direction for the coefficients $c_{ij}$, it is necessary to calculate the Jacobian. As this Jacobian contains the derivatives of the equations \pref{beqset} with respect to the coefficients\footnote{Note that the derivative of $\exp(c_{ij}T_j)$ w.r.t.\ the $c_{ij}$ is $T_j\exp(c_{ij}T_j)$. Hence no costly numerical derivatives of the dressing functions are necessary.} $c_{ij}$, each element contains essentially a two-dimensional integration. Calculating therefore each element one-by-one is extremely inefficient. It is therefore advantageous to evaluate all elements with fixed $p_j$ simultaneously for all derivatives, corresponding to a complete column in parallel. This parallelism is implemented by performing one addition of the Gauss-Legendre integration for each of the equations simultaneously by summing in an appropriate vector. Therefore it is also not possible to use standard Gauss-Legendre integrators from libraries, as they usually do not allow in a two dimensional array efficient parallel summation for a vector of integrals with appropriate treatment of input functions like the dressing functions.

The performance can be enhanced further by close inspection of the system under consideration. In the case of equations \prefr{fulleqG3d}{fulleqZ3d} it can be exploited that inside the integrals one of the dressing function does not depend on the integration angle. Excluding it from the angular integration also improves the cancellation of divergencies, thus reducing the required number of angular integration points. The same applies to factors of $T_j$ in the calculation of the Jacobian matrix.

Compared to the time necessary to calculate the Jacobian, all other elements are completely subleading. Therefore it is no advantage to use an approximate inversion routine to speed up the calculation of the inverse Jacobian for the Newton step. The gain in speed does not outweigh the drawback in precision. It is furthermore useful to store the inverted Jacobian and not to recalculate it during the backtracking, as it does not change. Using all these optimization gives a factor of roughly 30-60 in speed.

\subsection{Example calculation}\label{ssexample}

Here the results for an example calculation will be presented. The physical parameters are listed in table \ref{tabphpar}. The numerical parameters can be found in table \ref{tabnumpar} and the initial values along with the final values and run parameters in table \ref{tabinival}. The final values for the coefficients $c_{ij}$ are not displayed. Initially they have been set to 0. Using these parameters results have been obtained in a single super-cycle step.

\begin{table}
\begin{center}
\begin{tabular}{|c|c|c|c|c|c|c|c|}
\hline
$C_A$ & $g_3^2$ & $\delta_{3g}$ & $\zeta$ & $m_h$ & $m_r$ & $e_G$ & $e_Z$ \cr
\hline
3 & 1 & 1/4 & 1 & $0.8808g_3^2$ & $0.9930g_3^2$ & 0.39760 & -1.2952\cr
\hline
\end{tabular}
\end{center}
\caption{The physical parameters in the example run.}
\label{tabphpar}
\end{table}

\begin{table}
\begin{center}
\begin{tabular}{|c|c|c|c|c|c|c|c|c|}
\hline
$\delta$ & $\varepsilon$ & $\Lambda$ & $\Omega$ & $N_{ch}$ & $N_a$ & $N_\delta$ & $N_\varepsilon$ & $\alpha_G$ \cr
\hline
$10^{-6}g_3^2$ & $0.0201193g_3^2$ & $99407.2g_3^2$ & $10^7g_3^2$ & 39 & 80 & 60 & 100 & $10^{-5}$ \cr
\hline
$\sigma_1$ & $\sigma_0$ & $\delta_G$ & $N_N$ & $d$ & $\delta_I$ & $N_p$ & $N_\Lambda$ & \cr
\hline
0.5 & 0.1 & $10^{-8}$ & 5 & 2 & $10^{-5}$ & 100 & 60 & \cr
\hline
\end{tabular}
\end{center}
\caption{The numerical parameters in the example run.}
\label{tabnumpar}
\end{table}

\begin{table}
\begin{center}
\begin{tabular}{|c|c|c|c|}
\hline
Initial $A_G$ & 1.06894 & Final $A_G$ & 0.838007 \cr
\hline
Initial $A_Z$ & 0.99109 & Final $A_Z$ & 13.4111 \cr
\hline
Initial $A_H$ & 1.01411 & Final $A_H$ & 1.01404 \cr
\hline
Initial ratio $A_G^2 A_Z/J(\zeta,e_G)$ & 0.12035 & Final ratio $A_G^2 A_Z/J(\zeta,e_G)$ & 1.0019 \cr
\hline
Initial ratio $A_h^2/m_r^2$ & 1 & Final ratio $A_h^2/m_r^2$ & 0.99993 \cr
\hline
Initial $|\vec E|$ & $3.6\times 10^{5}$ & Final $|\vec E|$ & $2.5\times 10^{-9}$ \cr
\hline
Final rate of change of $A_G$ & $7.9\times 10^{-6}$ & Number of micro-cycles & 25 \cr
\hline
\end{tabular}
\end{center}
\caption{The initial and final values for the infrared coefficients together with some run statistics. The run time was about 20 minutes on a 2.8 GHz Intel Xeon processor.}
\label{tabinival}
\end{table}

\begin{figure}
\epsfig{file=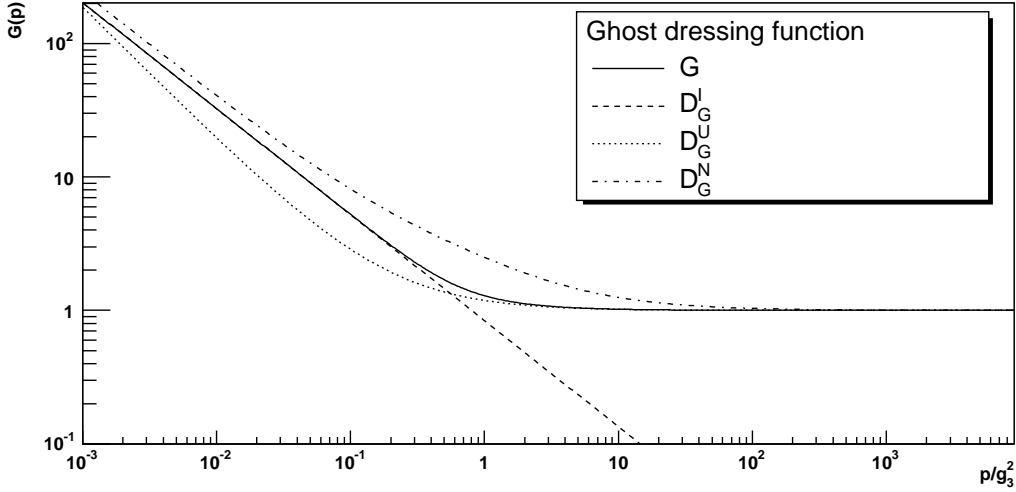,width=\linewidth}
\caption{The ghost dressing function compared to the infrared, intermediate, and ultraviolet fitting functions.}\label{figg}
\end{figure}

\begin{figure}
\epsfig{file=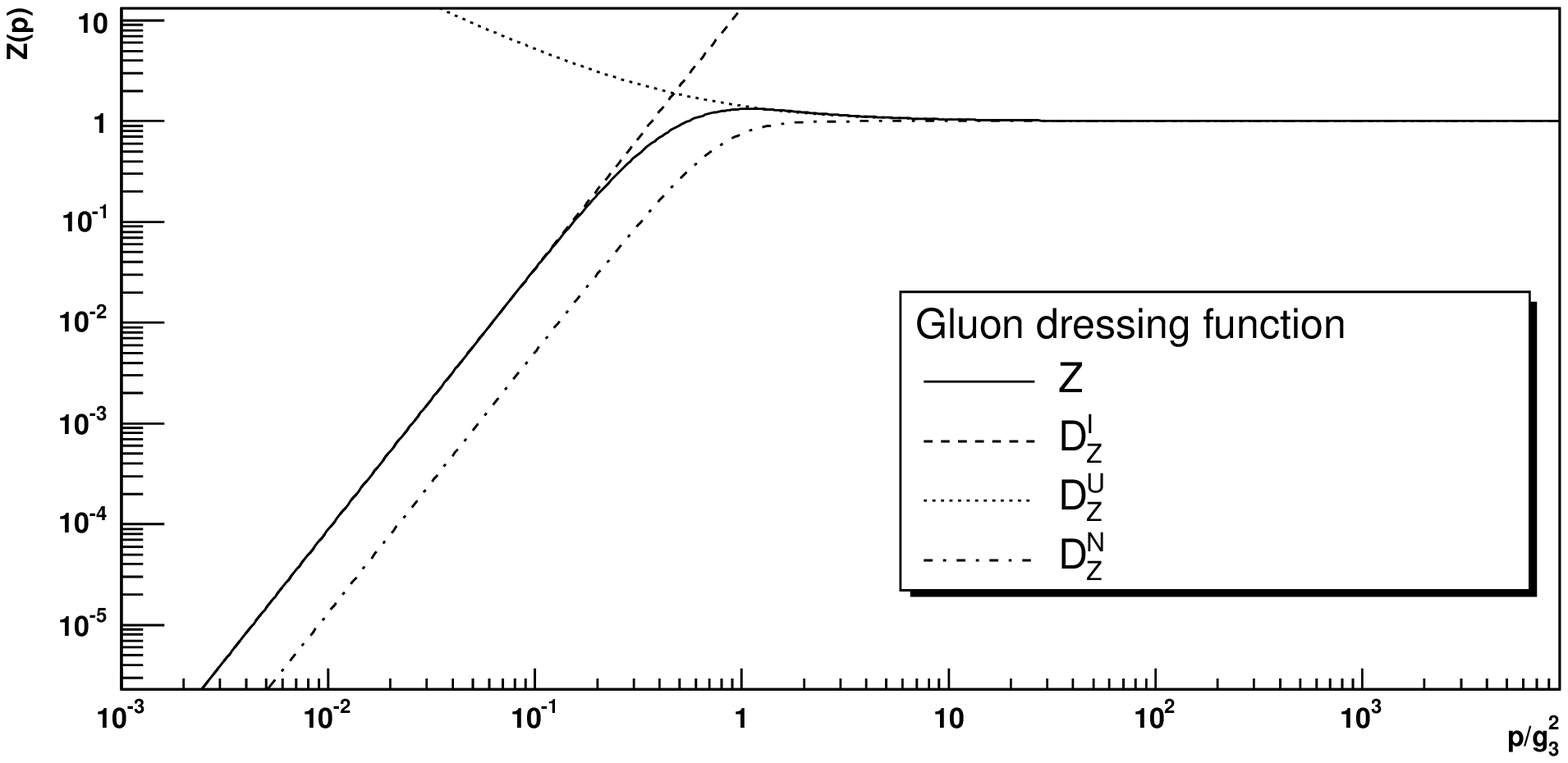,width=\linewidth}
\caption{The gluon dressing function compared to the infrared, intermediate, and ultraviolet fitting functions.}\label{figz}
\end{figure}

\begin{figure}
\epsfig{file=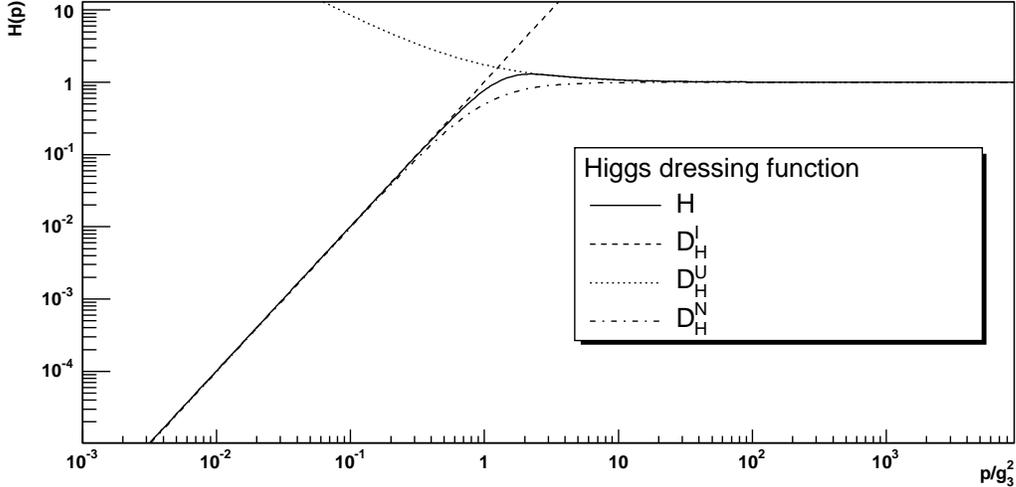,width=\linewidth}
\caption{The Higgs dressing function compared to the infrared, intermediate, and ultraviolet fitting functions.}\label{figh}
\end{figure}

The results are presented in figures \ref{figg}-\ref{figh} for the ghost, gluon, and Higgs dressing functions, respectively. Along the various final fitting functions in \pref{ffit} are shown. Together with the results in table \ref{tabinival} this demonstrates the abilities of the method in a most direct way. Note especially that the mismatch measured by $|\vec E|$ is decreased by 14 orders of magnitude. Up to 18 orders have also been achieved. Further examples can be found in refs.\ \cite{Maas:2005rf,Maas:2004se,Maas:2005hs}.

\section{4-dimensional Yang-Mills theory}\label{numericalren}

In case of the renormalized system \prefr{fulleqGft}{fulleqZft}, the main qualitative change is the appearance of the renormalization constants, defined in equations \prefr{ren1}{ren2}. Especially the infrared properties of the $p_0=0$ modes do not change \cite{Maas:2005rf,Maas:2005hs}. In the ultraviolet the coefficients in \pref{perturbative} change, but this effect can be ignored. In fact the subleading term is completely dropped and only the leading term is (and must be) retained. The $p_0\neq 0$ modes become constant both in the infrared and the ultraviolet.

If the renormalization constants would only appear explicitly as in equations \prefr{fulleqGft}{fulleqZft}, then the most direct implementation would be to treat them on equal footing with the remaining integrals. Although this would add $3N_{ch}$ integrals when evaluating the Jacobian, this would be a subleading effect. However, the constants cannot be kept the same during a complete Newton iteration. Although the system is finite, the effect of renormalization is still large and the would-be divergencies would lead to solutions which are incorrect, thus slowing the convergence significantly if not leading to divergence.

In addition, this is not the only effect. Due to the truncations made \cite{Maas:2005rf,Maas:2005hs}, it is necessary in nearly all integrals to replace the factor $F(q_0,\vec q)F(p_0\pm q_0,\vec p\pm \vec q)$ by\footnote{$Z_3$ is again the generic wave-function renormalization constant of the dressing function $F$.}
\be
\left(F(q_0,\vec q)-\frac{1}{Z_3(q_0)}\right)\left(F(p_0\pm q_0,\vec p\pm \vec q)-\frac{1}{Z_3(p_0\pm q_0)}\right).
\ee
Therefore the renormalization constants appear in a non-trivial combination, which are especially cumbersome when derivatives with respect to the $c_{ij}$ are performed. Although this problem could also be dealt with by determining the renormalization constants and their derivatives once for each $N_{ch}$ set of equations, a more direct approach is possible. For a single Newton step, it is viable to keep the renormalization constants fixed. Then afterwards new renormalization constants are determined and the next Newton step is performed. Therefore each Newton iteration is split into sub-iterations, each of one step length, which is repeated in a micro-cycle for $N_N$ steps or until convergence has been achieved. Due to the global convergence of the Newton method this is permissible and quite efficient, as the renormalization constants not change drastically within one Newton step. This strategy may not be possible in cases where these constants do change appreciably. In the present case, this is the most efficient solution. This is important, as now $3NN_{ch}$ equations have to be solved. With this method, a system of $N=20$ and $N_{ch}=60$ (yielding a Jacobian with about $10^7$ entries) converges without further prior knowledge within a few days on a 2.8 GHz Intel Xeon.

\section{Conclusions}\label{sconcl}

In this work a globally convergent method was presented to solve coupled sets of nonlinear integral equations. The example investigated are DSEs for Yang-Mills theories. Equipped with qualitative knowledge of the infrared and ultraviolet properties only, it was possible to solve the system without further prior knowledge with an acceptable efficiency. Most crucial to this efficiency is the careful optimization of the calculation of the Jacobian inside the global Newton method, which is at the heart of the algorithm.

The approach showed global convergence, thus extending earlier local algorithms, and therefore provides the abilities to solve such sets of equations without referring to other methods to provide good initial guesses. It will therefore be useful in the further investigations of DSEs and similar mathematical problems.

However, the method relies crucially on knowledge of the asymptotic behavior. Attempts to at least remove the need for the infrared behavior have failed so far, although many different methods have been tried, ranging from expansion in suitable characteristic functions up to genetic algorithms. In many physical cases these asymptotics are known, but especially in gauge theories this is usually not yet the case in general gauges, see e.g.\ \cite{Alkofer:2003jr}. Therefore it remains desirable to construct such a numerical continuum method.

Acknowledgments

The author is grateful to R. Alkofer, C. Feuchter, C. S. Fischer, B. Gr\"uter, and S. Roch for many helpful discussions and is indebted to R. Alkofer, C.~S.~Fischer and D. Nickel for a careful reading of the manuscript and helpful comments. This work is supported by the BMBF under grant number 06DA917 and 06DA116, and by the Helmholtz association (Virtual Theory Institute VH-VI-041).

\appendix

\section{Integral kernels and tadpoles of the 3-dimensional System}\label{appkernels}

The integral kernels in \prefr{fulleqG3d}{fulleqZ3d} are obtained using standard methods \cite{Alkofer:2000wg,Maas:2005rf,Maas:2004se}. They depend only on the magnitude $k$ and $q$ and the relative angle $\theta$ between their two momentum arguments. The kernel in the ghost equation \pref{fulleqG3d} is given by
\be
A_T(k,q)=-\frac{q^2\sin^3(\theta)}{(k^2+q^2-2kq\cos\theta)^2}.\nonumber
\ee
The contributions in the Higgs equation \pref{fulleqH3d} are
\bea
N_1(k,q)&=&-\frac{2q^2\sin^3(\theta)}{(k^2+q^2+2kq\cos\theta)^2}\\
N_2(k,q)&=&-\frac{2\sin^3(\theta)}{k^2+q^2+2kq\cos\theta}.
\eea
The kernels in the gluon equation \pref{fulleqZ3d} are finally
\bea
R(k,q)&=&-\frac{((\zeta-1)kq\cos(\theta)-q^2+\zeta q^2\cos^2(\theta))\nonumber
\sin\theta}{2k^2(k^2+q^2+2kq\cos\theta)}\\
M_L(k,q)&=&\frac{((\zeta-1)(k^2+4kq\cos\theta)-4q^2+4q^2\zeta\cos^2(\theta))\sin\theta}
{4k^2(k^2+q^2+2kq\cos\theta)}\\
M_T(k,q)&=&\frac{\sin\theta}{4k^2(k^2+q^2+2kq\cos\theta)}\Big((k^2+2q^2)((\zeta-9)k^2-4q^2)\nonumber\\
&&+8(\zeta-3)(k^2+q^2)kq\cos\theta+(8\zeta q^4+(\zeta+7)k^4\nonumber\\
&&+4(5\zeta-1)k^2q^2)\cos^2(\theta)+4(4\zeta q^2+(\zeta+3)k^2)\cos^3(\theta)\nonumber\\
&&+4\zeta k^2q^2\cos^4(\theta)\Big).
\eea
\no The modified gluon vertex is introduced by multiplying $M_T$ with
\be
(A(q,G,Z)A(q+k,G,Z)A(k,G,Z))^{\delta_{3g}}\label{g3vertex}
\ee
\no where $A(q,G,Z)$ is given by
\be
A(q,G,Z)=\frac{1}{Z(q)G(q)^{2+\frac{1}{2e_G}}}.
\ee
The tadpoles used read:
\bea
T^{GG}&=&-\frac{g_3^2C_A}{(2\pi)^2}\int dqd\theta\Big(R_D(p,q)(G(q)G(p+q)-A_G^2q^{-2e_G}(p+q)^{-2e_G})\nonumber\\
&&+R_3(p,q)\big(G(q)G(p+q)-A_G^2q^{-2e_G}(p+q)^{-2e_G}-1\big)\nonumber\\
&&+M_{TD}(p,q)Z(q)Z(p+q)\Big)\label{ggtad2}.
\eea
\noindent Here the kernel $R$ has been split into a convergent, $\zeta$-independent part $R_0$, its divergent part $R_D$, and a remaining part $R_3$ as
\be
R=R_0+(\zeta-3)R_3+R_D.\label{rsplit}
\ee
\noindent The divergent parts are in general isolated by
\be
R_D(p,q)=\lim_{q\to\infty}R(p,q).
\ee
\no $M_{TD}$ contains the divergent part of $M_T$, including the modifications due to the dressed 3-gluon-vertex \pref{g3vertex}. The other tadpole in the gluon equation is given by
\be
T^{GH}=-\frac{g^2C_A}{(2\pi)^2}\int dqd\theta\Big(M_{LD}(p,q)H(q)H(p+q)\Big).\label{ghtad}
\ee
\noindent Again, $M_{LD}$ contains the divergent part of $M_L$. In the Higgs equation, the tadpoles are set to
\be
T^{HG}+T^{HH}=\frac{g_3^2C_A}{p^2}\frac{m_h}{4\pi}.\label{higgstadpole}
\ee
\noindent A detailed account for the construction of the tadpoles is given in ref.\ \cite{Maas:2005rf}.

\end{document}